\documentclass{aa}
\usepackage{graphicx}
\usepackage{epsfig}
\usepackage{latexsym}
\usepackage{txfonts}
\usepackage{natbib}
\bibpunct{(}{)}{;}{a}{}{,}

\begin{document}
\renewcommand{\labelitemi}{-}
%title & authors
\title{Observational evidence for return currents in solar flare loops} \author{Marina Battaglia \and Arnold O. Benz}
\institute{Institute of Astronomy, ETH Zurich, 8092 Zurich,
Switzerland}
%\offprints{M. Battaglia, \email{battaglia@astro.phys.ethz.ch}}
\date{Received /Accepted}

%abstract
\abstract
%context
{ The common flare scenario comprises an acceleration site in the corona and particle transport to the chromosphere. Using satellites available to date it has become possible to distinguish between the two processes of acceleration and transport, and study the particle propagation in flare loops in detail, as well as complete comparisons with theoretical predictions.}
%aims
{We complete a quantitative comparison between flare hard X-ray spectra observed by RHESSI and theoretical predictions. This enables acceleration to be distinguished from transport and the nature of transport effects to be explored. }
%methods
{Data acquired by the RHESSI satellite were analyzed using full sun spectroscopy as well as imaging spectroscopy methods. Coronal source and footpoint spectra of well observed limb events were analyzed and quantitatively compared to theoretical predictions. New concepts are introduced to existing models to resolve discrepancies between observations and predictions. }
%results
{The standard thin-thick target solar flare model cannot explain the observations of all events. In the events presented here, propagation effects in the form of non-collisional energy loss are of importance to explain the observations. We demonstrate that those energy losses can be interpreted in terms of an electric field in the flare loop. One event seems consistent with particle propagation or acceleration in lower than average density in the coronal source.}
%conclusions
{We find observational evidence for an electric field in flare loops caused by return currents.}

\titlerunning{Return currents}
\authorrunning{Marina Battaglia \& Arnold O. Benz}

\maketitle

%----------------------------------------------------------------
% Introduction
%-----------------------------------------------------------------

%
\section{Introduction} \label{Introduction}

Solar flares have been studied in detail both observationally and theoretically ever since their discovery by \citet{Ca859}. Although increasingly more sophisticated instrumentation provides ever more detailed data, we still lack the basic understanding of many processes at work in a solar flare. 

The common flare picture as deduced from hard X-ray (HXR) observations features an HXR source in the corona~\citep[coronal or loop-top source,][]{Fr71,Hu78}, and two or more HXR sources (footpoints) in the chromosphere \citep{Hoy81}. These sources are thought to be due to bremsstrahlung emission produced by fast electrons accelerated somewhere above the loop. If we assume that a single particle beam creates both coronal and footpoint emission, the most basic model would involve thin target emission at the top of the coronal loop and thick target emission from the footpoints, which both produce characteristic spectra. 
\citet{Wh95} developed a more sophisticated model (intermediate thin-thick target or ITTT model) to fit observations by Yohkoh. They based their model on observations by \citet{Fe94}, who found high column densities at the loop top, which might act as a thick target below a certain electron energy. In the ITTT model, the shape of the coronal and footpoint non-thermal spectra and the relation between them, observed by Yohkoh, can be explained. The column density in the coronal source determines a critical energy level for the electrons. Electrons that have an energy below this critical energy are stopped in the coronal region. Consequently, the distribution of electron energies measured at the footpoints is depleted in low energy electrons. If the column density is high, the coronal source may act as a thick target to electrons of energies as high as 60 keV, which would leave almost no footpoint emission. Observational evidence for such coronal thick targets were found in RHESSI observations \citep[eg.][]{Ve04}.

Less extreme cases, flares with one or more footpoints, have frequently been observed by RHESSI.
To study the spectral time evolution of individual sources, five well-observed events were analyzed by \citet{Ba06} who focused on the differences between the spectral indices of coronal and footpoint spectra. They found that in two of those events, the differences at specific times as well as the time-averaged difference was significantly larger than two, ruling out a simple thin-thick target interpretation. In \citet{Ba07}, the spectra of the five events were compared with the predictions of the ITTT model. The authors exploited the order of magnitude improvement in spectral resolution of RHESSI over the 4-point Yohkoh spectra and showed that most RHESSI observations could not be explained by the ITTT model. 

\citet{Ba07} proposed that by considering non-collisional energy loss inside the loop this inconsistency could be resolved.  
A possible mechanism that causes non-collisional energy loss is an electric field. Accelerating electrons out of the coronal source region drives a return current to maintain charge neutrality in the whole loop. For finite conductivity, Ohm's law implies that an electric field must be present. The beam electrons lose energy because of work expended in moving inside the electric potential. This produces a change in the shape of the electron spectrum at the footpoints. The formation and evolution of these return currents were studied by various authors \cite[e.g.][]{Kn77,SS84,La89,Oo90}. \citet{Zh06} proposed that return currents could explain the high energy break observed in flare HXR spectra. 
Most studies have, however, been theoretical proposals or numerical simulations, based on standard flare values, that do not attempt to explain or reproduce true solar flare observations. 

\citet{Ba07} compared RHESSI spectra to the ITTT-model of \citet{Wh95}, demonstrating that the qualitative shape and relations between coronal and footpoint-spectrum often do not agree with the model predictions. In this study, we take an additional step by completing a quantitative analysis of the relation linking coronal and footpoint spectra in the context of the thin-thick target model; we demonstrate that, in some cases, electric fields related to return currents can indeed explain the relation between coronal and footpoint spectra.

In Sect.~\ref{Theory} we summarize the basic physical concepts applied in the paper. Section~\ref{evdescription} provides a brief overview of the analyzed events and a description of the spectral analysis. In Sect. \ref{Method}, we describe our calculation of the energy loss required to reproduce the observed footpoint spectrum, constrained by the coronal emission. Our results are presented in Sect.~\ref{results}. In Sect. \ref{retcurrnefield}, we link those results to the concept of return currents. 

%----------------------------------------------------------------------------------
%Theory
%----------------------------------------------------------------------------------
\section{Thin and Thick target emission} \label{Theory}

Two types of bremsstrahlung emission are distinguished. If the electrons pass a target without losing a significant amount of energy, the corresponding emission is referred to as thin target \citep{Da73}. This situation is expected to occur in coronal regions when electrons pass through a target of insufficient column density to stop them. If the electrons are fully stopped inside the target, the resultant emission is called thick target emission \citep{Br71}. This is the case for the dense chromospheric material at the footpoints. 

For an input power-law electron distribution of the shape $F(E)=A_{E}E^{-\delta}$, the non-relativistic bremsstrahlung theory predicts power-law photon spectra 
\[I(\epsilon)\propto \epsilon^{-\gamma}\quad\mbox{where}\quad\left\{\begin{array}{l} \gamma=\delta+1\quad\mbox{in the thin target case} \\ \gamma=\delta-1\quad\mbox{in the thick target case} \end{array} \right. \] 
The observable distinction between the two emission mechanisms is a difference $\Delta \gamma$ of value 2 in their observed photon spectral indices. 

Assuming a final column density in the coronal source, the coronal source spectrum is a thick target at low energies and a thin target at high energies with a break at some critical energy. The footpoint spectrum is depleted at low energies, as low energy electrons do not reach the chromosphere. An illustration of this can be found in \citet{Wh95} or \citet{Ba07}.

In the events that we analyze here, a thermal component is present in all observations. Observation of the non-thermal emission is therefore only possible at photon energies higher than 15 keV.  For these energies we can assume that the coronal source is a pure thin target and the footpoints are a pure thick target.

%---------------------------------------------------------------------------------------
%Event description and data analysis
%---------------------------------------------------------------------------------------

\section{Event description and spectral analysis} \label{evdescription}
\begin{figure*} 
\resizebox{\hsize}{!}{\includegraphics{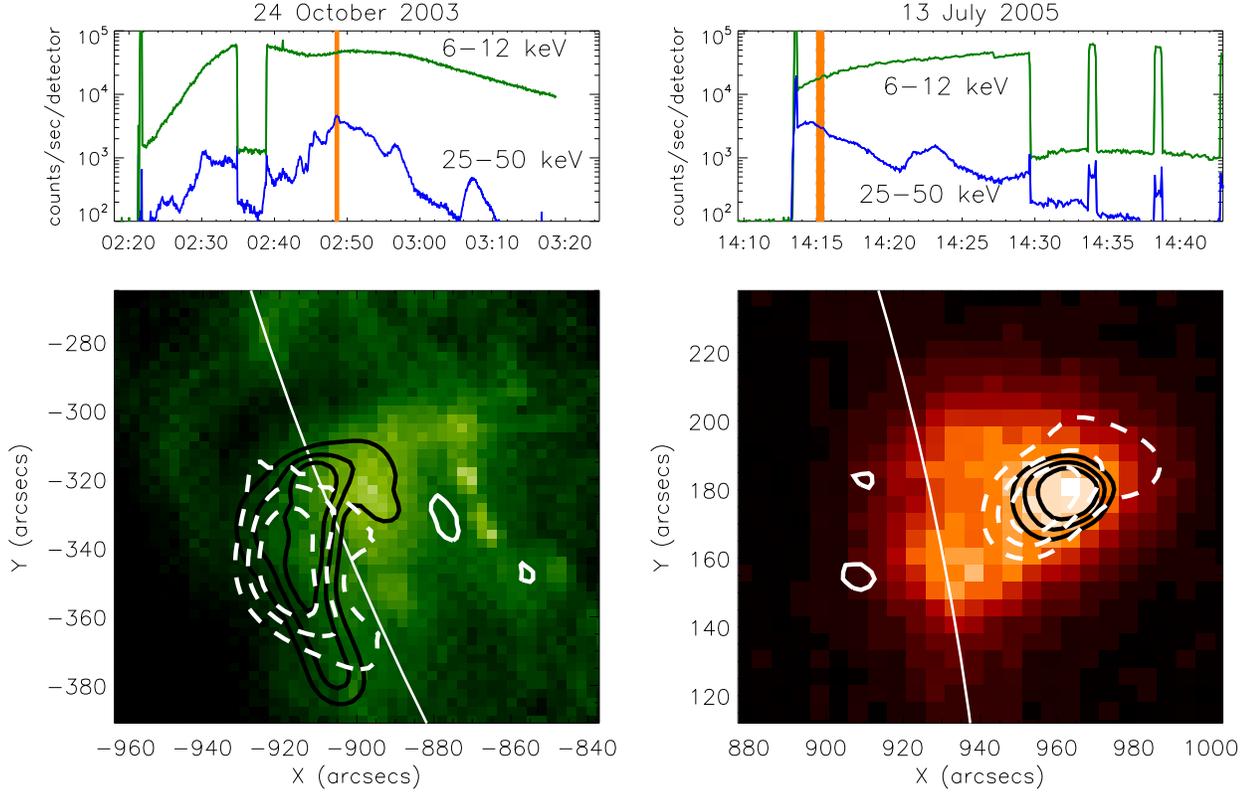}}
\caption {\textit{Top}: RHESSI light curves in the 6-12 and 25-50 keV energy band. The analyzed time interval is indicated by the \textit{vertical bar}. \textit{Bottom}: SOHO/EIT image at 24 October 2003 02:47:31 (\textit{left}), GOES SXI image at 13 July 2005 14:19:04 (\textit{right}). The 30 \%, 50 \% and 70 \% contours from RHESSI Pixon images of the coronal source at 12-16 keV (\textit{solid contours}) and 20-24 keV (\textit{dashed contours}, mainly non-thermal emission) are overlaid along with the 50\% contour of the footpoint sources at 25-50 keV.}
\label{2flimages}
\end{figure*}

The two events that we analyze were described in detail by \citet{Ba06, Ba07}. They were selected because of the significant differences between the footpoint and coronal non-thermal spectral index.  The first event occurred on 24 October 2003 around 02:00 UT (GOES M7.7), the second on 13 July 2005 around 14:15 UT (GOES M5.1). Both events occurred close to the limb but were not occulted; two footpoints were therefore fully observed in both cases. RHESSI light curves from the time of main emission are shown in Fig.~\ref{2flimages}. An EIT image of the 24 October 2003 event and a GOES SXI image for the event of 13 July 2005 are presented for orientation. The contours of the coronal source and the footpoints from RHESSI images are overlaid. 

\subsection{Spectral fitting and analysis} \label{fitting}

The two events were analyzed using imaging spectroscopy with the PIXON  algorithm \citep{Me96, Hur02}. Images were made in a 30 second time interval during which the flux was sufficiently high for good images but pile-up was low. The image times are given in Table~\ref{fitpartable} and shown in Fig.~\ref{2flimages}. The spectra of the footpoints and the coronal source were measured and fitted. The regions of interest from which the spectra were computed were chosen to be a circle around the coronal source and a polygon around the footpoints to include all of the emission at all energies. The effects of this method of region selection are discussed in \cite{Ba05}. As a simplification, the footpoints were treated as one region and the spectrum was fitted with a single power-law. In the presence of the footpoints, the non-thermal emission in coronal sources is difficult to observe. Therefore, two methods of fitting the coronal source were used. First, a thermal component was fitted to the spectrum at low energies and a single power-law to the energies higher than about 25 keV. As a second method, the full sun thermal spectrum was fitted. As shown in \cite{Ba05}, the thermal emission observed in full sun spectra is mostly coronal emission. We therefore used the thermal full sun fit as an approximation to the coronal thermal component and completed a power-law fit at the higher energies, while the thermal emission was fixed. This supports the idea that non-thermal emission exists in the coronal source and provides an estimate of the accuracy of the non-thermal coronal fit. The energy ranges for the fits were 8-36 keV for the coronal source and 24-80 keV for the footpoints. All fit parameters are provided in Table~\ref{fitpartable}.

In the thin-thick target model an electron beam is assumed to be injected into the center of the coronal source.
The column depth that the electrons travel through in the corona is then $\Delta N=n_e\cdot l$, where the path length \textit{l} is half the coronal source length. 
From RHESSI images in the 10-12 keV band, the source area \textit{A} was measured to be the 50\% contour of the maximum emission. We approximate the source volume to be $V=A^{3/2}$ and the path length to be $l=\sqrt{A}/2$. Using the observed emission measure EM, the particle density is computed to be $n_e=\sqrt{EM/V}$ which corresponds to a column depth of $\Delta N=\sqrt{EM}A^{-1/4}/2$ expressed in observable terms.
A volume filling factor of  1 was assumed for the computation of the density, which will be improved in Sect.~\ref{expfemissionloss}.
The emission measures were taken from the spectral fits to the coronal source and to full sun spectra. Additionally, temperatures and emission measures observed by GOES were included. This provides a range for the emission measures, temperatures, and column depths, and an estimate of their uncertainty.  \\

%----------------------------------------------------------------------------------------
%Method
%----------------------------------------------------------------------------------------
\section{Method}\label{Method}

\begin{figure*} 
\resizebox{\hsize}{!}{\includegraphics{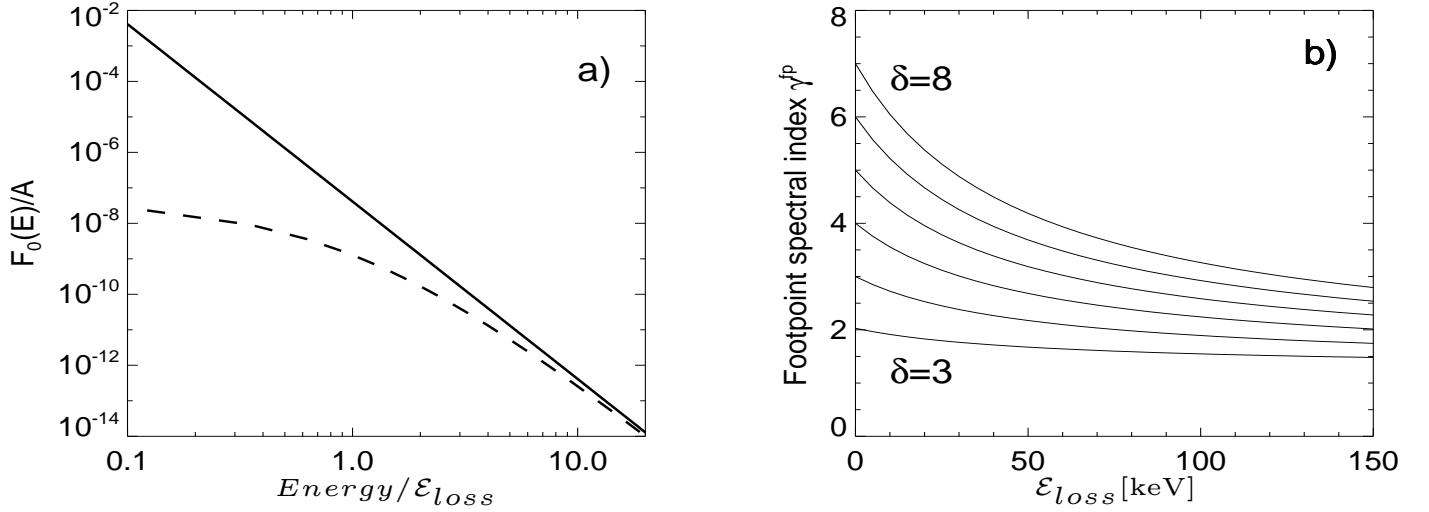}}
\caption {a) Normalized electron power-law spectrum (\textit{solid}) and altered spectrum due to a constant energy loss of 30 keV (\textit{dashed}). b) Relation between loss energy $\mathcal{E}_{loss}$ and fitted thick target photon power-law spectral index $\gamma^{fp}$ for initial electron spectral index $\delta=3,4,5,6,7,8$ in the accelerator.}
\label{elsp}
\end{figure*}

Starting from the assumption that the observed coronal spectrum at high photon energies is caused by thin target emission, we compute the electron distribution and therefore the expected footpoint photon spectrum. 
This is completed via the following steps.
\begin{enumerate}
\item We assume that the observed coronal photon spectrum can be fitted by a power law:
\begin{equation}
 F^{cs}_{obs}(\epsilon)=A_{\epsilon}^{cs}\epsilon^{-\gamma^{cs}},
\end{equation} where A$_{\epsilon}^{cs}$ is the normalization and $\gamma^{cs}$ the photon spectral index.
\item Using thin target emission, the injected electron spectrum F(E) is then proportional to  \label{itemelsp}
\begin{equation} \label{bla}
F(E)=A_E E^{-\delta}\sim\frac{A_{\epsilon}^{cs}}{\Delta N}E^{-\gamma^{cs}+1}
\end{equation} where $\Delta N$ is the column depth the electrons travel through inside the coronal target \citep{Da73}. 
\item The expected thick target emission $F_{exp}^{fp}(\epsilon)$ caused by this electron distribution in the footpoints can be computed as follows \citep{Br71}:
\begin{equation}
 F^{fp}_{exp}(\epsilon)=A_{\epsilon,exp}^{fp}\epsilon^{-\gamma^{fp}_{exp}}\sim\frac{A_{\epsilon}^{cs}}{\Delta N}\epsilon^{-(\gamma^{cs}-2)}
\end{equation}
The superscripts fp and cs denote the footpoint and coronal source values, respectively.
\item The normalization and spectral index of $F^{fp}_{exp}(\epsilon)$ is compared to the observed footpoint spectrum $F^{fp}_{obs}(\epsilon)$.
\end{enumerate}
In thin-thick target models, the difference in spectral index $\Delta \gamma=|\gamma^{cs}-\gamma^{fp}|$ is 2. As the observed difference is larger than 2 in the selected events, a mechanism has to be found that causes the electron spectrum to harden while the beam passes down the loop. We present a mechanism that assumes an electric field which causes electrons to lose the energy $\mathcal{E}_{loss}$ independently of the initial electron energy. The resulting spectrum is flatter, although not strictly a power-law function anymore (Fig.~\ref{elsp}a). The deviation becomes substantial below 2 $\mathcal{E}_{loss}$.

The necessary energy loss is determined as follows:
\begin{enumerate}
\item We start with the coronal electron distribution as found from Point~\ref{itemelsp} in the above list.
\item By assuming a thin target, the electron distribution leaves the coronal source and propagates down the loop. A constant energy loss $\mathcal{E}_{loss}$ is subtracted from the electron energies as the energy loss is independent of the electron energy.
\item We compute the expected thick target photon spectrum $F^{fp}_{exp}(\epsilon)$ from this altered electron spectrum.
\item A power law is fitted to  $F^{fp}_{exp}(\epsilon)$. The fitted energy range is 30-80~keV. This is the range for which footpoint emission is typically observed. 
\end{enumerate}
The relation between the energy loss experienced  and the corresponding photon spectral index of the best-fit power law function depends on the initial electron spectral index $\delta$ and the energy loss $\mathcal{E}_{loss}$. It is equivalent to the elementary charge times the electric potential between the coronal source and the footpoints. If the initial electron spectral index is 8 for instance, the thick target photon spectral index without energy loss is 7. With increasing energy loss, this value decreases rapidly. The effect is less pronounced when the initial electron spectrum is harder. This is shown in Fig.~\ref{elsp}b for several values of $\delta$ and $\mathcal{E}_{loss}$. Using the curves in this figure, we can easily determine the energy loss that causes an electron spectrum of spectral index $\delta$ to result in a fitted photon spectral index $\gamma^{fp}$. 

% The coronal source was fitted with a thermal component and a non-thermal power-law component. The footpoints were treated as one region and fitted with one power-law. From this first analysis, the events could be divided into three subsets of events.
%\begin{itemize}
%\item Events with difference in spectral index between footpoints and coronal source  $\Delta \gamma > 2$ (Events 3,8)
%\item Events with $\Delta \gamma = 2$ (Events 1,6,7)
%\item Events with $\Delta \gamma < 2$ (Events 2,4,5,9)
%\end{itemize}
%The events with $\Delta \gamma < 2$ are not well separated and the difference can be attributed to an observed mixed thin-thick target. Events 3 and 8 are used to extend the thin-thick target model. Fig.\ref{2flimages} shows EIT images of the two events with RHESSI contours overlaid.

\begin{table*}

%\begin{minipage}[t][10cm]{\columnwidth}
\begin{minipage}[t][]{2\columnwidth}

\caption{Overview of main event properties and fit parameters }
\renewcommand{\footnoterule}{}  % to avoid a line before footnotes
\begin{tabular}{l|lll|lll}   % c = center, l = left justified, r=right just.
\hline \hline
Time interval &\multicolumn{3}{c}{24 October 2003 02:48:20-02:48:50 }& \multicolumn{3}{c}{13 July 2005 14:15:00-14:15:30 }\\
\hline
Area [cm$^2$] / Volume [cm$^3$]&\multicolumn{3}{c}{$7.9\cdot 10^{18}$ / $2.2\cdot 10^{28}$}&\multicolumn{3}{c}{$1.7\cdot 10^{18}$ / $2.2\cdot 10^{27}$} \\
\hline
&full sun&imspec& GOES&full sun&imspec&GOES\\
\hline
Temperature [MK]\footnote{ Thermal parameters for three different measuring methods (RHESSI full sun fit, RHESSI imaging spectroscopy, GOES). See also comment in Sect~\ref{obsspectra}. }&21.6&23.2&15.4&23.8&22.3&18.1\\
Emission measure [$10^{49}$cm$^{-3}$]&0.98&0.44&3.4&0.22&0.22&0.47\\
Electron density [$10^{10}$cm$^{-3}$]&2.1&1.4&3.9&3.2&3.2&4.6\\
Column density [$10^{19}$cm$^{-2}$]&2.9&2.0&5.5&2.1&2.1&3.0 \\
\hline
&footpoints&cs fit 1&cs fit2&footpoints&cs fit 1&cs fit 2\\
$\gamma$\footnote{ Two different values (cs fit 1, cs fit 2) for the non-thermal coronal fit, distinguishing the two fitting methods used (compare Sect~\ref{fitting}).}&2.6&6.2&6.1&2.9&5.1&5.6 \\
$F_{50}[\mathrm{photons\, cm^{-2}s^{-1}keV^{-1}}]$&2.0&0.07&0.07&0.88&0.06&0.05 \\

\hline
\end{tabular}
\label{fitpartable}
\end{minipage}

\end{table*}

%--------------------------------------------------------------------------------------
%Results
%--------------------------------------------------------------------------------------

\section{Results}\label{results}

\subsection{Observed spectra} \label{obsspectra}

\begin{figure*} 
\resizebox{\hsize}{!}{\includegraphics{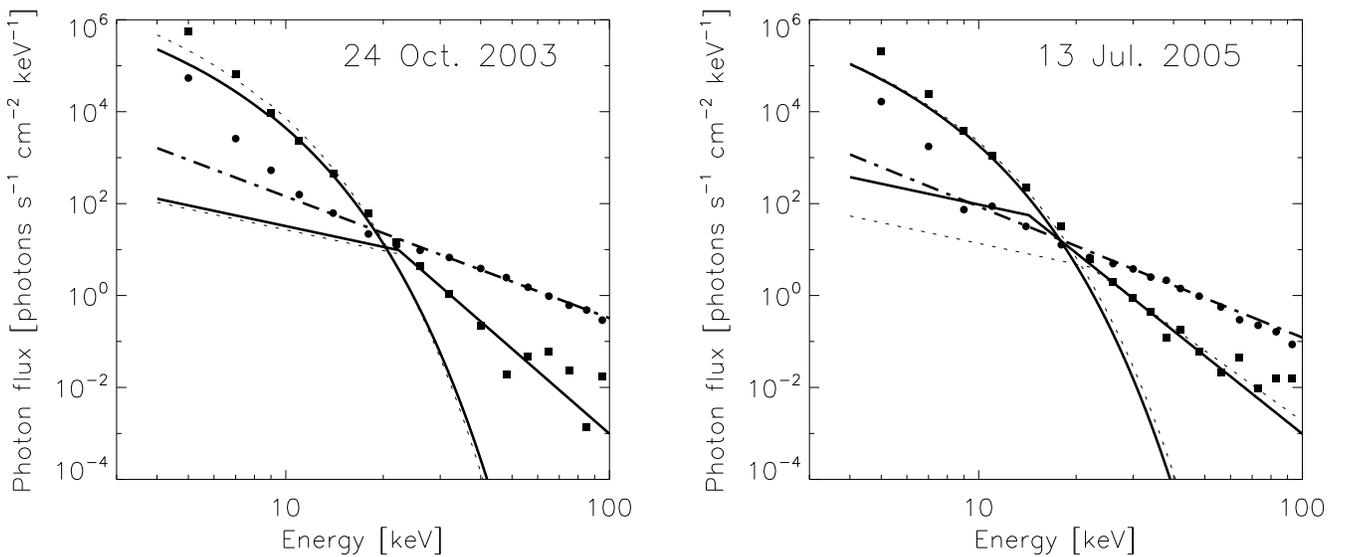}}
\caption {Coronal and footpoint source spectra overlaid with the according fits. \textit{Dots} are the measured footpoint spectrum, the \textit{dashed-dotted} line indicates the fit to this spectrum in the range 30-80 keV. \textit{Squares} indicate the observed coronal source spectrum. The \textit{solid} lines provide the thermal and non-thermal fits as found from imaging spectroscopy. The \textit{dotted} lines give the thermal fit to the full sun spectrum and the resulting non-thermal fit in imaging spectroscopy.}
\label{spectra}
\end{figure*}

Figure~\ref{spectra} shows the observed spectra overlaid with the spectral fits. As indicated in Sect. \ref{fitting}, the thermal fits differ slightly from each-other. The main reason why the fits do not agree is the wider energy binning adopted by imaging spectroscopy. With this binning, the atomic lines are not resolved, contrary to full sun spectroscopy.

Using the different fitting methods as an estimate of the uncertainty, an average difference in spectral index of $\Delta \gamma = 3.55 \pm 0.07$ for the event of 24 October 2003 and $\Delta \gamma = 2.45 \pm 0.35$ for the event of 13 July 2005 is found between the coronal source and footpoints. 
\subsection{Expected footpoint emission and energy loss} \label{expfemissionloss}
\begin{figure*} 
\resizebox{\hsize}{!}{\includegraphics{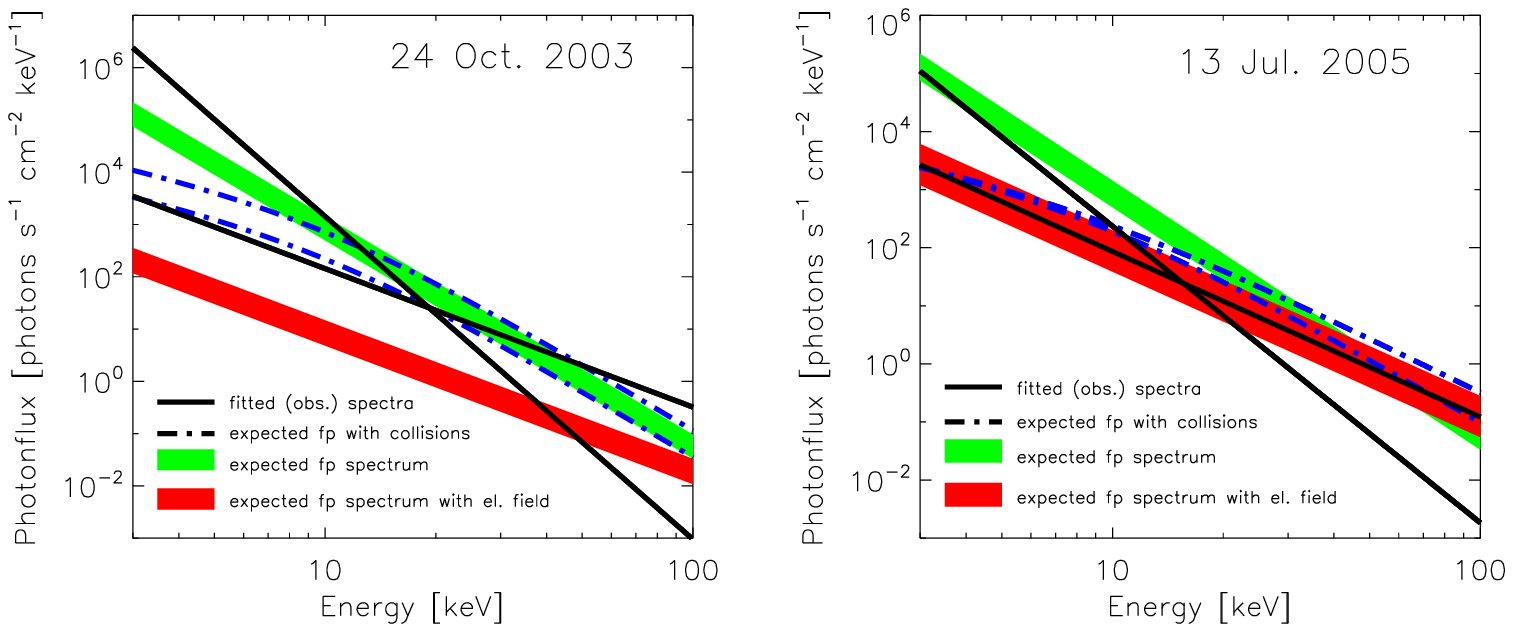}}
\caption {Observed (fitted) power-laws of the non-thermal coronal source and footpoints (\textit{solid}). The \textit{light-gray (green)} area indicates the range of expected footpoint spectra without energy loss. The \textit{dark-gray (red)} area marks the range of expected footpoint spectra when energy loss is applied to the electrons to find the same spectral index as the observed footpoint spectrum. The \textit{dash-dotted} lines give the expected spectra at the footpoints if the only transport effect was Coulomb collisions of the beam electrons (cf. Sect.~\ref{collisions}).}
\label{efieldspec}
\end{figure*}

As described in Sect.~\ref{Method}, we computed the electron distribution from the coronal source photon spectrum and the expected thick target emission (footpoint spectrum) caused by this electron distribution. Figure~\ref{efieldspec} shows the measured spectra, the expected footpoint spectrum from a pure thick target, and the footpoint spectrum when introducing energy loss. 

To compute the electron flux, the coronal column density is required. As given in Table~\ref{fitpartable}, three different values of the column density were estimated. Using those, we are able to reproduce a range of possible electron spectra (Table \ref{beamvalues}) and, therefore footpoint spectra. The confidence range of the footpoint spectra is indicated by a light gray (green) area in Fig.~\ref{efieldspec}.

The derived energy loss depends on the fitted coronal and footpoint spectra. For the two different coronal fitting methods, a range of loss energies $\mathcal{E}_{loss}=[58.0,59.4]$ keV  is found for the event of 24 October 2003 and $\mathcal{E}_{loss}=[8.7,26]$ keV for the event of 13 July 2005. The normalization of the new spectrum depends on the initial electron distribution. The initial electron distribution is computed according to Eq.~(\ref{bla}). It depends on the column depth. If the column depth is lower, more electrons are needed to produce the same X-ray intensity. The resulting range of possible spectra is shaded in dark-gray in Fig.~\ref{efieldspec}. As shown in the figure, the footpoint spectrum with energy loss reproduces well the observed footpoint spectrum for the event of 13 July 2005. 

During the event of 24 October 2003, the predicted footpoint spectrum is, however, an order of magnitude less intense than observed. In the context of the ITTT model, this implies that the electron flux density emanating from the coronal region is higher than predicted. The discrepancy may be explained by density inhomogeneities in the coronal source resulting in a smaller effective column density. The electron flux is underestimated if the non-thermal X-ray emission originates in regions that are less dense than average. In the following we therefore assume that the coronal source has an inhomogeneous density; this is represented by dense regions with filling factor smaller than 1 for thermal emission and, for the non-thermal coronal source in the 24 October 2003 event, a density that is lower by an order of magnitude. The observations do not allow to determine the filling factor.
As can be seen from Fig.~\ref{2flimages}, the coronal source in the event of 13 July 2005 is very compact, while the source in the event of 24 October 2003 is more extended, showing isolated intense regions. This supports the assumption that the density is inhomogeneous in the 24 October 2003 coronal source and the true column density in the X-ray emitting plasma might be smaller than deduced from the measurements. Since $F_{exp}^{fp}(\epsilon)$ is proportional to $1/\Delta N$, an effective column density of an order of magnitude less than the observed could produce the observed footpoint spectrum. 
In the computations presented in Sect.~\ref{retcurrnefield}, we assume an effective column density $\Delta N_{eff} = \Delta N/14$ for the event of 24 October 2003.

\section{Return current and electric field} \label{retcurrnefield}
In the above analysis, we assumed that the electrons experienced a constant energy loss while streaming down the loop. We now demonstrate that this energy loss could be caused by an electric potential in the loop that drives a return current. 
There was much controversy surrounding the precise physical mechanism that generates the return current \citep[eg.][]{Kn77, SS84, Oo90}. The basic scenario is the following: We assume that the electrons are accelerated in the coronal source region. When a beam of accelerated electrons, which is not balanced by an equal beam of ions, leaves this region, a return current prohibits charge build-up and the induction of a beam-associated magnetic field. In the return current, thermal electrons move towards the coronal source. Since their velocity is relatively small, they collide with background ions and cause resistivity. Ohm's law then implies the presence of an electric field in the downward direction.  
The return current density $j_{ret}$ can be derived from the equation of motion for the background electrons \citep{Bebook}.
\begin{equation}
\frac{\partial \vec{v}}{\partial t}+(\vec{v}\cdot \triangledown)\vec{v}=-\frac{e}{m}\vec{E_{ind}}-\frac{e}{mc}(\vec{v}\times \vec{B})-\nu_{e,i}\vec{v}
\end{equation}
where $\vec{E_{ind}}$ is the electric field induced by the return current, $\vec{v}$ is the mean velocity of the electrons that represent the return current, and $\nu_{e,i}$ is the electron-ion collision frequency. Using $\vec{j_{ret}}=-en\vec{v}$, this expression can be written as
\begin{equation} \label{fullohm}
\left(\frac{\partial}{\partial t}+\nu_{e,i}\right)\vec{j_{ret}}=\frac{(\omega_p^e)^2}{4\pi}\vec{E_{ind}}+\frac{e}{mc}[\vec{j_{ret}}\times (\vec{B_0}+\vec{B_{ind}})],
\end{equation}
where $B_0$ is the guiding magnetic field and $B_{ind}$ the field induced by the beam. We neglect the last term on the right side of Eq.~\ref{fullohm} by assuming that the beam and return currents are anti-parallel, oriented along the guiding magnetic field, and that the perpendicular component of $B_{ind}$ vanishes. We then obtain
\begin{equation}
\left(\frac{\partial}{\partial t}+\nu_{e,i}\right)\vec{j_{ret}}=\frac{(\omega_p^e)^2}{4\pi}\vec{E_{ind}}.
\end{equation}

Since we consider a fixed time interval that is far longer than the collision time, we assume a steady state, neglecting the time derivative of the return current. The equation then takes the form of the classical Ohm's law:
\begin{equation}
\vec{j_{ret}}=\frac{(\omega_p^e)^2}{4\pi \nu_{e,i}}\vec{E_{ind}}=\sigma \vec{E_{ind}}.
\end{equation}

We estimate whether the energy loss computed in Sect.~\ref{results} is caused by this electric field.
From the observations, we estimated the electron loss energy $\mathcal{E}_{loss}$ that the electrons experience in the loop (Sect.~\ref{expfemissionloss}). Assuming this loss is caused by the induced electric field $E_{ind}$ and across the distance \textit{s} from the coronal source to the footpoints (i.e. half the loop length), we compute the electric field to be
\begin{equation}
E_{ind}=\frac{\mathcal{E}_{loss}}{e\cdot s}.
\end{equation}

Using Spitzer conductivity~\citep{Spbook}, the term for the return current is related to the observed loss energy by:
\begin{equation} \label{returneq}
j_{ret}=6.9\cdot 10^6T_{loop}^{3/2}\frac{\mathcal{E}_{loss}}{e\cdot s}\quad \mathrm{[statamp/cm^2]},
\end{equation}
where $T_{loop}$ is the temperature in the loop.

On the other hand, the beam current density can be written as
\begin{equation} \label{jbeam}
j_{beam}=\frac{F_{tot}(E)}{A_{fp}}\cdot e \quad \mathrm{[statamp/cm^2]},
\end{equation}
where $A_{fp}$ is the total footpoint area. The total electron flux per second $F^{tot}(E)$ is computed from the observed electron spectrum as follows: Let the electron spectrum be $F(E)=A_eE^{-\delta}$. The total flux of streaming electrons per second above a cutoff energy $E_{cut}$ is then:
\begin{equation} 
F_{tot}(E)=\int_{E_{cut}}^{\infty}F(E) \mathrm{d}E=\frac{A_e}{\delta-1}E_{cut}^{-(\delta-1)}.
\end{equation}

In a steady state, the relation 
\begin{equation}
j_{beam}=j_{ret}
\end{equation}
 is valid. \\

Comparing the beam current as described in Eq.~(\ref{jbeam}) with the return current from the observed energy loss according to Eq.~(\ref{returneq}), we test whether the assumption of Spitzer conductivity holds.

\subsection{Results} \label{retresults}
Table \ref{beamvalues} presents the relevant physical parameters necessary for the derivation of the beam- and return currents.
For Spitzer conductivity, the loop temperature $T_{loop}$ is required. Its is expected to have a value between the coronal source temperature and the footpoint temperature (see Table~\ref{beamvalues}). As a first assumption, a mean temperature of $T_{loop}=15$ MK is chosen. The loop length is evaluated from RHESSI images, approximating the distance between the sources from the centroid positions and assuming a symmetrical loop structure. This provides a typical half loop length of $4\cdot 10^9$ cm. 
The footpoint area is measured from the 50\% contour in RHESSI images in the 25-50 keV energy range, yielding a total footpoint area of $\approx (6-7)\cdot 10^{17}$ cm$^2$. 
The beam current density depends critically on the electron cut off energy $E_{cut}$. We use a value of 20 keV. This gives an approximate lower limit to the total amount of streaming electrons.  

Using the presented observations, Eq.(\ref{jbeam}) and by assuming Spitzer conductivity (Eq. \ref{returneq}), the return current results to be of an order of magnitude higher than the beam current. This contradicts the assumptions of a steady state, and is also unphysical. 
\begin{table*}
\caption{Values used for the computation of the beam and return currents and computed currents. } 
\begin{tabular}{lrr}   % c = center, l = left justified, r=right just.
\hline \hline
Event & 24 October 2003 & 13 July 2005 \\
\hline
Assumed loop temperature $T_{loop}$ [MK] & 15 & 15 \\
1/2 Loop length s [cm] & $3.2\cdot 10^9$&$4.3\cdot 10^9$     \\
Electron flux $F(E)$ [s$^{-1}$ keV$^{-1}$]&6.33$\cdot 10^{41}E^{-5.2}$-1.5$\cdot 10d^{42}E^{-5.2}$&1.7$\cdot10^{40}E^{-4.1}-8.4\cdot 10^{40}E^{-4.6}$\\
Electron cutoff energy [keV] & 20 & 20  \\
Total footpoint area [cm$^2$] & $7.2\cdot10^{17}$& $6.2\cdot10^{17}$ \\
$E_{loss}$ [keV] & 58-59.4 & 8.7-26 \\
Electric field strength [statvolt/cm] &$(6-6.3) \cdot 10^{-8}$&$(6.7-20.3)\cdot 10^{-9}$ \\
\hline
$j_{ret}$[$statamp/cm^2$]&$(2.4-2.5) \cdot 10^{10}$& $(2.7-8.1)\cdot 10^9$\\
$j_{beam}$ [$statamp/cm^2$]&$(5.1-14.4) \cdot 10^9$&$(1.9-3.6) \cdot 10^8$ \\
\hline
\end{tabular}

\label{beamvalues}
\end{table*}

\section{Discussion} \label{discussion}

\subsection{Instability}
In Sect.~\ref{retresults}, we assumed Spitzer conductivity when computing the return current which produced the unphysical result of $j_{ret}>j_{beam}$. Using Eq.~(\ref{returneq}), the loop temperature required to maintain equality between the return current and beam current ($j_{ret}=j_{beam}$) can be computed. In the 24 October 2003 event, the loop temperature $T_{loop}$ would need to be smaller than 10 MK; in the 13 July 2005 event, $T_{loop}$ should be less than 3.9 MK. Such low loop temperatures are highly unlikely.
However, it is possible that the return current is unstable to wave growth. For an extended discussion of instabilities in parallel electric currents, see e.g. \citet{Bebook}. Instability causes an enhanced effective collision frequency of electrons in the return current and therefore a lower effective conductivity. The ion cyclotron instability develops if the drift velocity of the beam particles $V_d$ exceeds the thermal ion velocity $v_{th}^{ion}$ as follows: 

\begin{equation} \label{instcond}
V_d\ge15\frac{T_i}{T_e}v_{th}^{ion}
\end{equation}
with $v_{th}^{ion}=\sqrt{\frac{k_BT_i}{m_i}}$ and $T_e$ and $T_i$ being the electron and ion temperatures, respectively. 

We assume a steady state for which $j_{beam}=j_{ret}=n_e eV_d$, where $V_d$ is the mean drift velocity of the electrons constituting the return current. We therefore express $V_d$ as
\begin{equation}
V_d=\frac{j_{beam}}{n_e e}
\end{equation}
and substitute this expression and that for $v_{th}^{ion}$ in Eq. (\ref{instcond}). Assuming $T_e = T_i=T_{loop}$ and solving Eq.~(\ref{instcond}) for $T_{loop}$ the instability condition holds
\begin{equation} \label{tcond}
T_{loop}\le 2.3\cdot 10^8 \left(\frac{j_{beam}}{n_e}\right)^2 \quad \mathrm{[K]}.
\end{equation}

\begin{figure} 
\resizebox{\hsize}{!}{\includegraphics{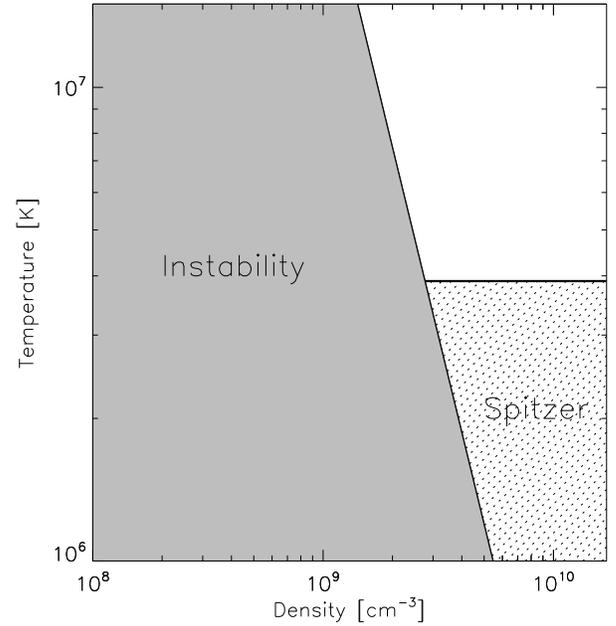}}
\caption {Region of instability in the density/temperature space for the event of 13 July 2005. The \textit{grey} region indicates the densities and temperatures $T_{loop}$ for which the return current is unstable (Eq.~\ref{instcond}). In the lower right part, the return current is stable and $j_{beam}=j_{ret}$ with Spitzer conductivity. The current in the event of 24 October 2003 is unstable for all values of density and temperature in the Figure.}
\label{instab}
\end{figure}

Since the loop temperature and density are not known exactly, 
this relation is illustrated in Fig.~\ref{instab} for several values of $n_e$ and $T_{loop}$ typical in flare loops. For the values of $j_{beam}$ found for the observations of the two events, we find that instability occurs in the 24 October 2003 event for all values of $n_e$ and $T_{loop}$ in Fig.~\ref{instab}.
For the 13 July 2005 flare, three distinct regions in the diagram can be found. The solid line indicates the relation of Eq. (\ref{tcond}). Below this line, the return current is unstable. At high densities and low temperatures (lower right), the return current is stable and $j_{beam}=j_{ret}$ with Spitzer conductivity. The range of beam currents $j_{beam}$ deduced from the data allows for loop temperatures $< 3.9$ MK. 
In the upper right quadrangle, Spitzer conductivity would imply $j_{ret}>j_{beam}$, which is unphysical. If the loop was in this parameter range, the current instability would be most likely saturated and $T_e>T_i$. This would shift the instability threshold in Fig.~\ref{instab} to the right. Further, a loop in the state presented by the uppermost part of the figure (temperature above 10 MK, high density) would be detectable by the RHESSI satellite even in the presence of the coronal source. Since no loop emission is observed, we conclude that the loop is either less dense, cooler or both. The values in the upper right quadrangle are therefore unlikely.

\subsection{Low energy electron cutoff}
In the above computations, a value of 20 keV for the electron cutoff energy $E_{cut}$ was assumed. This value is within the range for which the thermal and non-thermal components of the spectrum intersect. Values around 20 keV or higher are also supported by detailed studies of the exact determination of low-energy cutoffs \citep[eg.][]{Sa05,Ve05,Sui07}. What if the cutoff energy were substantially lower than 20 keV? A cutoff energy as low as 10 keV would increase the total electron flux and therefore the beam current by an order of magnitude, leading to $j_{ret}\approx j_{beam}$ for Spitzer conductivity. Conductivity could then not be reduced significantly by wave turbulence, and instability would be marginal. If the low energy cutoff were even lower than 10 keV, we would find that $j_{ret}< j_{beam}$. This could not be explained in terms of the model used here.
\subsection{Source inhomogeneity and filling factor} \label{filling}
In the above paragraphs, it was demonstrated that the energy loss for the electrons due to an electric field could resolve the inconsistency in the difference between footpoint and coronal source spectral indices (Sect. \ref{results}). For the event of the 24 October 2003, this produces footpoint emission that is lower than the observed emission (Fig. \ref{efieldspec}). Images show that the coronal source in this event is not compact, but extended with brighter and darker regions. It is therefore possible that the standard density estimate, which favors high density regions, produces higher densities than average and that the effective column density of the regions, where the largest part of the non-thermal emission originates, is lower. This would provide a higher expected footpoint emission, in closer agreement with observations. 

As mentioned in Sect.~\ref{fitting}, a filling factor of 1 was used for the computation of the column density. A filling factor smaller than 1 would lead to even higher densities of the SXR emitting plasma. However, this would not affect the lower effective column density of the regions, where the HXR emission originates when assuming source inhomogeneity. 

\subsection{Collisions and other possible scenarios} \label{collisions}
While return currents may not be the only means of attaining non-collisional energy loss, they are the most obvious and best studied. However, other scenarios are conceivable, which could produce a harder footpoint spectrum (or a softer coronal spectrum). 
In the model presented here, collisional energy loss of beam electrons is neglected. This is a valid assumption for the following reasons: If collisions of the beam electrons in the loop were to play an important role, significant HXR emission should originate in the loop. Within the dynamical range limitations of RHESSI, this is not the case. However, GOES SXI and SOHO/EIT images imply that the loop is filled with hot material. To study possible effects of collisions, we compared the change in the electron spectrum and the resulting footpoint spectrum for collisional energy loss and energy loss due to the electric field. The change in the electron spectrum due to collisions depends on the column depth through which the beam passes and was computed by \citet{Le81} and \citet{Br75}. We assume a column depth derived from the density in the loop times half the loop length. Assuming the same density as in the coronal source, we derive an upper limit to the collisional effects. The expected footpoint spectrum from purely collisional losses is indicated in Fig.~\ref{efieldspec} as dash-dotted lines. Collisions affect the low energetic electrons most where a significant change in the spectrum is found. At the higher energies observed in this study, the spectrum does not change significantly. The neglect of collisional effects is therefore justified.
\citet{Br08} showed that in certain cases, emission from non-thermal recombination can be important, generating a coronal spectrum that is steeper than expected by the thin-target model. This could also produce a difference in the spectral index that is larger than two. Acceleration over an extended region \citep[as proposed by][]{Xu08} could alter the electron distribution at the footpoints. If the distribution was harder at the edge of the region, a spectral index difference larger than 2 would result. A thorough comparison of such models with observations may be the scope of future work. 
\section{Conclusions} \label{conclusions}
The spectral relations between coronal and footpoint HXR-sources provide information about electron transport processes in the coronal loop between the coronal source and the footpoints. Most models neglect these processes in the prediction of the shape and quantitative differences between the source spectra. As shown by \citet{Ba07}, the observations of some solar flares do not fit the predictions of such models, in particular the intermediate thin-thick target model by \citet{Wh95}: there is a discrepancy concerning the difference in coronal and footpoint spectral indices, which is expected to be 2. 

We have analyzed the two out of five events that display a spectral index difference larger than two in more detail. Such a behavior can be attributed to energy loss during transport that is not proportional to electron energy, but $\mathcal{E}_{loss}/E$ is larger at low energies. Such an energy loss causes the footpoint spectrum to flatten, which increases the difference in spectral indices. Two loss mechanisms come to mind immediately: Coulomb collisions and an electric potential. Figure 4 demonstrates that the assumption of an electric potential reproduces the observations more accurately.

In one of the two events, there remains a discrepancy between the observed and expected footpoint emission, such that the electron flux at the footpoints is larger than predicted. This flux was estimated from the observed non-thermal HXR (photon) flux and the observed thermal emission of the coronal source. We attribute the discrepancy to propagation or acceleration in low density plasma, which also heats the adjacent high-density regions. 

The energy loss can therefore be explained by an electric field in the loop associated to the return current, which builds up as a reaction to the electrons streaming down the loop and the associated beam current. In a steady state ($j_{beam}=j_{ret}$), the return current is unstable to wave growth in one event for all realistic temperature and density parameters in the loop. The kinetic current instability drives a wave turbulence that enhances the electric resistivity by many orders of magnitude. This anomalous resistivity in turn significantly enhances the electric field. In the event of 13 July 2005, the return current may be stable if the loop density is high and the temperature is low, and Spitzer conductivity is applied. Both cases (out of five) present strong evidence for a return current in flares for the first time.

Transport effects by return currents constitute a considerable energy input by Ohmic heating into the loop outside the acceleration region. It may be observable in EUV. Comprehensive MHD modeling including the coronal source, the footpoints, and the region in-between, may be the goal of future theoretical work.

\begin{acknowledgements}
RHESSI data analysis at ETH Z\"urich is supported by ETH grant TH-1/04-2 and the Swiss National Science Foundation (grant 20-105366). 
We thank S\"am Krucker for helpful comments and discussions.
\end{acknowledgements}

\bibliographystyle{aa}
\bibliography{mybib}

\end{document}